# Wormholes in Einstein-Born-Infeld theory

Martín G. Richarte* and Claudio Simeone†

*Departamento de Física, Facultad de Ciencias Exactas y Naturales, Universidad de Buenos Aires, Ciudad Universitaria, Pabellón I, 1428 Buenos Aires, Argentina*

Spherically symmetric thin-shell wormholes are studied within the framework of Einstein-Born-Infeld theory. We analyze the exotic matter content, and find that for certain values of the Born-Infeld parameter the amount of exotic matter on the shell can be reduced in relation with the Maxwell case. We also examine the mechanical stability of the wormhole configurations under radial perturbations preserving the spherical symmetry.

## I. INTRODUCTION

Traversable Lorentzian wormholes [1,2] are topologically nontrivial solutions of the equations of gravity, which would imply a connection between two regions of the same universe, or of two universes, by a traversable throat. In the case that such geometries actually exist they could show some interesting peculiarities as, for example, the possibility of using them for time travel [3,4]. A basic difficulty with wormholes is that the flareout condition [5] to be satisfied at the throat requires the presence of matter that violates the energy conditions ("exotic matter") [1,2,5,6]. It was recently shown [7], however, that the amount of exotic matter necessary for supporting a wormhole geometry can be made infinitesimally small. Thus, in subsequent works special attention has been devoted to quantifying the amount of exotic matter [8,9], and this measure of the exoticity has been pointed as an indicator of the physical viability of a traversable wormhole [10]. Theories beyond the Einstein-Maxwell framework have been explored with interesting results in this sense [11].

A central aspect of any solution of the equations of gravitation is its mechanical stability. The stability of wormholes has been thoroughly studied for the case of small perturbations preserving the original symmetry of the configurations. In particular, Poisson and Visser [12] developed a straightforward approach for analyzing this aspect for thin-shell wormholes, that is, those which are mathematically constructed by cutting and pasting two manifolds to obtain a new manifold [13]. In these wormholes the associated supporting matter is located on a shell placed at the joining surface; so the theoretical tools for treating them is the Darmois-Israel formalism, which leads to the Lanczos equations [14,15]. The solution of the Lanczos equations gives the dynamical evolution of the wormhole once an equation of state for the matter on the shell is provided. Such a procedure has been subsequently followed to study the stability of more general spherically symmetric configurations (see, for example, Ref. [16]), and an analogous analysis has also been carried out in the case of cylindrical symmetry (see [17]).

In order to avoid an infinite energy density associated to the electron, Born and Infeld [18] introduced a fundamental field strength $b$, which together with the electron charge $e$ determines a characteristic length $r_0 = \sqrt{e/b}$. The associated electromagnetic theory is nonlinear, and among its consequences one can find a regular gravitational field for a fundamental electrically charged particle. From a theoretical point of view, this feature would be enough to suggest a revision of the well-known gravitational effects of charged objects within the wider framework of nonlinear electrodynamics. Besides, Born-Infeld (BI)-type actions were later recovered in the context of low energy string theory [19]. For these reasons, in the last years renewed attention has been devoted to spherically symmetric gravitational fields in the framework of Einstein gravity coupled to Born-Infeld or other nonlinear electrodynamics [20,21].

---

*martin@df.uba.ar

†csimeone@df.uba.ar

In particular, wormholes within the framework of Einstein gravity and nonlinear electrodynamics were considered, for example, in Ref. [22]. In the present work, we study wormholes of the thin-shell type associated to spherically symmetric solutions of general relativity and the Born-Infeld theory of electromagnetism. In Sec. II, we introduce Einstein-Born-Infeld (EBI) spherically symmetric geometry, and starting from it, in Sec. III, we mathematically construct the associated thin-shell wormholes. In Sec. IV, we evaluate the amount of exotic matter required for the existence of the wormholes, and in Sec. V, we study the mechanical stability under perturbations preserving the spherical symmetry. Throughout the paper we use natural units, so $G = c = 1$.

## II. THE THEORY

Let us begin with a review of the main characteristics of the spherically symmetric solutions found in the framework of EBI theory. For a four-dimensional manifold $(\mathcal{M}_4, g_{\mu\nu})$ with nonvanishing cosmological constant $\Lambda$, the action takes the following form:

$$S = \int_{\mathcal{M}_4} d^4x \sqrt{|g_4|} \left( \frac{1}{2\kappa^2} (R - 2\Lambda) + \mathcal{L}(F) \right),$$ (1)

where $g_4 = \det g_{\mu\nu}$, and $\mathcal{L}(F)$ represents the Born-Infeld Lagrangian

$$\mathcal{L}(F) = 4b^2 \left( 1 - \sqrt{1 + \frac{F_{\mu\nu} F^{\mu\nu}}{2b^2}} \right),$$ (2)

with $b$ a parameter which has dimension of length or mass, the so-called Born-Infeld parameter. By taking the limit $b \to \infty$, $\mathcal{L}(F)$ reduces to the Maxwell Lagrangian

$$\mathcal{L}(F) = -F_{\mu\nu} F^{\mu\nu} + \mathcal{O}(F^4).$$ (3)

By varying the action above with respect to the gauge field $A_\mu$ and the metric tensor $g_{\mu\nu}$ the field equations for the spacetime metric and the electromagnetic field are obtained:

$$\nabla_\mu \left( \frac{F^{\mu\nu}}{\sqrt{1 + \frac{F^2}{2b^2}}} \right) = 0,$$ (4)

$$G_{\mu\nu} + \Lambda g_{\mu\nu} = T^{\text{BI}}_{\mu\nu},$$ (5)

$$T^{\text{BI}}_{\mu\nu} = \frac{1}{2} \mathcal{L}(F) g_{\mu\nu} + \frac{2 F_{\kappa\mu} F^\kappa_\nu}{\sqrt{1 + \frac{F^2}{2b^2}}}.$$ (6)

It has been proven that for any $d$-dimensional static spherically symmetric spacetime the metric can be written in terms of the hypergeometric function [23]. In the particular case of four-dimensional spacetime the metric reads

$$ds^2 = -g(r) dt^2 + \frac{1}{g(r)} dr^2 + r^2 d\Omega_2^2,$$ (7)

where the function $g(r)$ takes the form

$$g(r) = 1 - \frac{2m}{r} - \frac{\Lambda r^2}{3} + \frac{2b^2 r^2}{3} \left( 1 - \sqrt{1 + \frac{q^2}{b^2 r^4}} \right) + \frac{4q^2}{3r^2} F_1 \left[ \frac{1}{4}, \frac{1}{2}, \frac{5}{4}, -\frac{q^2}{b^2 r^4} \right].$$ (8)

Here, $m$ is an integration constant related with the Arnowitt-Deser-Misner mass of the configuration, and $q$ is the charge of the system. The solution above (with nonzero cosmological constant) was found by Fernando and Krug [24]. For large $r$ the metric represents a correction to the Reissner-Nordström (RN) anti-de Sitter black hole:

$$g(r) = 1 - \frac{2m}{r} + \frac{q^2}{r^2} - \frac{\Lambda r^2}{3} - \frac{q^4}{40b^2 r^6}.$$ (9)

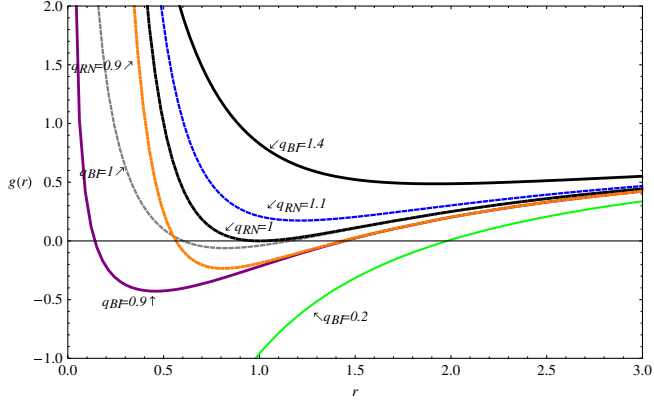

FIG. 1 (color online). We show the position of the horizons for the RN geometry ($b \to \infty$), with $m = 1$, $q_{\rm RN} = 0.9$, 1, 1.1 and the BI spacetime in the case of $m = 1 = b$ and $q_{\rm BI} = 0.2$, 0.9, 1, 1.4. We show how the position of the horizons changes with the parameter by varying the charge in both metrics.

In the case $\Lambda = 0$ and in the limit $b \to \infty$ this metric reduces to the standard RN geometry. The last term in (9) represents the first Born-Infeld correction to the Reissner-Nordström anti-de Sitter black hole in the large $b$ limit. Instead, near the origin ($r = 0$) the metric presents a completely different behavior when compared with the Reissner-Nordström geometry; for small $r$ we have

$$g(r) = 1 - \frac{2m - a}{r} + 2b\left(-q + \frac{br^2}{3} + \frac{b^2 r^4}{10}\right), \tag{10}$$

$$a^2 = q^3 \frac{b}{\pi} \Gamma^4\left[\frac{1}{4}\right], \tag{11}$$

that is, close to the origin the leading term in the metric is given by $(2m - a)/r$. In fact, when $2m = a$ the metric is smoothed at $r = 0$, so the nonlinear Born-Infeld source helps to regularize the metric at the position of the charge. The same analysis can be carried out for the only nonzero component of the stress tensor, obtaining that the electric field is also finite at $r = 0$. This fact is not surprising because the Born-Infeld theory was developed in order to have a finite self-energy associated to a pointlike charge.

Besides, by demanding $g(r = r_{\rm hor}) = 0$ we can obtain the mass in terms of the horizon position $r_{\rm hor}$ (see Fig. 1). In order to keep the main ideas as clear as possible we only show one case in which we contrast the extremal RN geometry with the BI one. For a fixed value of $m$ and $b$, by increasing the parameter $q_{\rm BI}$ in the range [0.2, 1.4] we could have one, two, or zero horizons (see Fig. 1). If $m$ and $q_{\rm BI}$ are fixed, and the BI parameter $b$ varies, we arrive at the same result. In conclusion, EBI black holes seem to be quite interesting because they include a wide variety of geometries.

## III. THIN-SHELL WORMHOLE CONSTRUCTION

Starting from the metric given by (7) we build a spherically symmetric thin-shell wormhole in the Einstein-Born-Infeld theory. We take two copies of the spacetime and remove from each manifold the four-dimensional regions described by

$$\mathcal{R}_{1,2} = \{x / r_{1,2} \le a, a > r_{\rm hor}\}. \tag{12}$$

The resulting manifolds have boundaries given by the timelike hypersurfaces

$$\Sigma_{1,2} = \{x / r_{1,2} = a, a > r_{\rm hor}\}. \tag{13}$$

Then we identify these two timelike hypersurfaces to obtain a geodesically complete new manifold $\mathcal{M}$ with a matter shell at the surface $r = a$, where the throat of the wormhole is located. This manifold is constituted by two regions, which in the case $\Lambda = 0$ are asymptotically flat (see Fig. 2). To study this type of wormhole we apply the Darmois-Israel formalism [14] to the case of Einstein-Born-Infeld theory. We can introduce the coordinates $\xi^i = (\tau, \theta, \chi)$ in $\Sigma$, with $\tau$ the proper time on the throat. Though we will first focus in static configurations, in the subsequent analysis of the mechanical stability of the configuration, we must allow the radius of the throat to be a function of the proper time; then in general we have that the boundary hypersurface reads

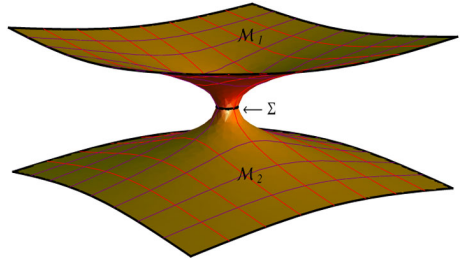

FIG. 2 (color online). We show the wormhole geometry obtained after performing the cut and paste procedure. The shell on $\Sigma$ is located at the throat radius $r = a$.

$$\Sigma: \ \mathcal{F}(r, \tau) = r - a(\tau) = 0. \tag{14}$$

The field equations projected on the shell $\Sigma$ are obtained within the Darmois-Israel formalism; this leads to the Lanczos equations [14]

$$\langle K_{ab} \rangle - \langle K \rangle h_{ab} = -\kappa^2 S_{ab}, \tag{15}$$

where the bracket $\langle . \rangle$ stands for the jump of a given quantity across the hypersurface $\Sigma$. The tensor $h_{ab}$ is the induced metric on $\Sigma$, and the extrinsic curvature tensor $K_{ab}$ is defined as follows:

$$K_{ab}^{\pm} = -n_c^{\pm} \left( \frac{\partial^2 X^c}{\partial \xi^a \partial \xi^b} + \Gamma_{de}^c \frac{\partial X^d}{\partial \xi^a} \frac{\partial X^e}{\partial \xi^b} \right)_{r=a}, \tag{16}$$

where $n_c^{\pm}$ are the unit normals ($n_c n^c = 1$) to the surface $\Sigma$. After some algebraic manipulations, the nonzero components $S_a^b$ of the surface energy-momentun tensor of the shell turn to be given by

$$S_\tau^\tau = \frac{1}{2\pi a} \left( \sqrt{\dot{a}^2 + g(a)} \right), \tag{17}$$

$$S_\phi^\phi = S_\theta^\theta = \frac{1}{8\pi a} \left( \frac{2\dot{a}^2 + a g'(a) + a\ddot{a} + 2g(a)}{\sqrt{\dot{a}^2 + g(a)}} \right), \tag{18}$$

where the dot means a derivative with respect to the proper time and the prime with respect to $a$. From these equations we read the energy density $\sigma = -S_\tau^\tau$ and the transverse pressure $p = S_\theta^\theta = S_\phi^\phi$ in terms of the throat radius $a(\tau)$, first and second derivatives of $a(\tau)$, and the function $g(a)$, which depends on the parameters of the system. If we explicitly write $g(r)$ and take the limit $b \to \infty$ in both Eqs. (17) and (18), we recover the expression for the energy density $\sigma$ and pressure $p$ found in the work by Eiroa and Romero (see Ref. [16]). with the Lanczos equations for the case associated to the standard RN solution.

It is easy to see from Eqs. (17) and (18) that the energy conservation equation is fulfilled:

$$\frac{d(A\sigma)}{d\tau} + p \frac{dA}{d\tau} = 0, \tag{19}$$

where $A$ is the area of the wormhole throat. The first term in Eq. (19) represents the internal energy change of the shell, and the second work by internal forces of the shell. The dynamical evolution of the wormhole throat is governed by the Lanczos equations and to close the system we must supply an equation of state $p = p(\sigma)$ that relates $p$ and $\sigma$.

In the next section we will study how the exotic matter amount is related to the BI parameter, that is, we will mainly analyze the differences between the amount of exotic matter in a Born-Infeld wormhole and in a RN one.

## IV. AMOUNT OF EXOTIC MATTER

Motivated by the nonlinear structure of BI theory, we will evaluate the amount of exotic matter and the energy conditions. Essentially, we want to know if it is possible that for certain values of the BI parameter $b$ the amount of exotic matter located at the shell can be reduced in relation to the Maxwell case, that is, if this quantity is larger or smaller than the corresponding to a RN wormhole.

The weak energy condition (WEC) states that for any timelike vector $U^\mu$ it must be $T_{\mu\nu}U^\mu U^\nu \geq 0$; the WEC also implies, by continuity, the null energy condition, which means that for any null vector $k^\mu$ it must be $T_{\mu\nu}k^\mu k^\nu \geq 0$ [2]. In an orthonormal basis the WEC reads $\rho \geq 0$, $\rho + p_l \geq 0 \ \forall l$, while the null energy condition takes the form $\rho + p_l \geq 0 \ \forall l$. In the case of thin-shell wormholes the radial pressure $p_r$ of matter on the shell is zero, and within the relativistic theory of gravitation the surface energy density must fulfill $\sigma < 0$; thus, both energy conditions would be violated. The sign of $\sigma + p_t$ where $p_t$ is the transverse pressure is not fixed, but it depends on the values of the parameters of the system. In this section we restrict to static configurations. The surface energy density $\sigma_0$ and the transverse pressure $p_0$ for a static configuration ($a = a_0$, $\dot{a}_0 = 0$, $\ddot{a}_0 = 0$) are given by

$$\sigma_0 = -\frac{1}{2\pi a_0}\sqrt{g(a_0)}, \tag{20}$$

$$p_0 = \frac{1}{4\pi a_0}\left(\frac{a_0 g'(a_0) + g(a_0)}{\sqrt{g(a_0)}}\right). \tag{21}$$

The most usual choice for quantifying the amount of exotic matter in a Lorentzian wormhole is the integral [9]

$$\Omega = \int(\rho + p_r)\sqrt{|g_4|}d^3x. \tag{22}$$

We can introduce a new radial coordinate $R = \pm(r - a_0)$, with $\pm$ corresponding to each side of the shell. Then, because in our construction the energy density is located on the surface, we can also write $\rho = \delta(R)\sigma_0$, and because the shell does not exert radial pressure the amount of exotic matter reads

$$\Omega = \int_0^{2\pi}\int_0^\pi\int_{-\infty}^{+\infty}\delta(R)\sigma_0\sqrt{|g_4|}dR\sin\theta d\theta d\phi = 4\pi a_0^2\sigma_0. \tag{23}$$

Replacing the explicit form of $\sigma_0$ and $g_4$, we obtain the exotic matter amount as a function of the parameters that

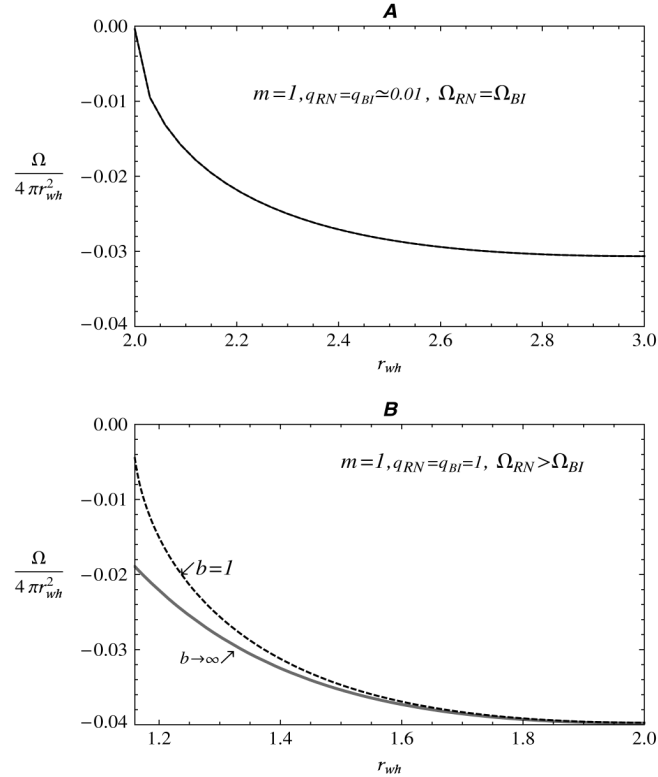

FIG. 3. We show the energy densities corresponding to wormholes associated to the RN geometry (solid line) and the BI spacetime (dashed line) in terms of the wormhole radius $r_{\text{wh}}$. Here, we take $r_{\text{wh}} > r_{\text{hor}}^+$ with $r_{\text{hor}}^+$, indicating the largest event horizon of both the RN and BI cases. The absolute value of the energy density is smaller in the BI case.

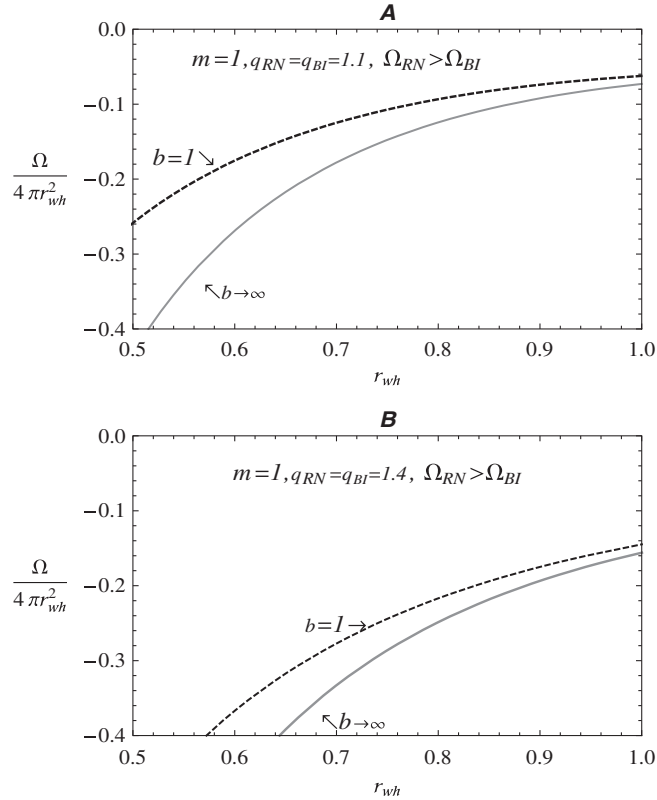

FIG. 4. The energy densities corresponding to the BI wormholes in terms of $r_{wh}$ are exhibited for $q \in [1.1, 1.4]$ in the case $m = b = 1$. As before, the absolute value of the energy density is smaller in the BI case.

characterize the configurations

$$\Omega = -2a_0 \left( 1 - \frac{2m}{a_0} - \frac{\Lambda a_0^2}{3} + \frac{2b^2 a_0^2}{3} - \frac{2b^2 a_0^2}{3} \left( 1 + \frac{q^2}{b^2 a_0^4} \right)^{1/2} + \frac{4q^2}{3a_0^2} F_1 \left[ \frac{1}{4}, \frac{1}{2}, \frac{5}{4}, -\frac{q^2}{b^2 a_0^4} \right] \right)^{1/2}.$$

In the case $\Lambda = 0$ and in the limit $b \to \infty$, we obtain the amount of exotic matter for the wormholes associated to the Reissner-Nordström ($q \neq 0$) and Schwarzschild ($q = 0$) geometries.

We always take the throat radius larger than the event horizon radius (when horizons exist in the original manifold). For small charge values ($q \approx 0.01$), the RN geometry as well as the BI geometry (with $b = 1$) show the same behavior (see Fig. 3(a)). However, by comparing the extremal RN geometry with $q = 1$ and the BI geometry with the same charge (with $b = 1$) the amount of exotic matter turns out to be smaller for the BI geometry. So, even when both geometries have the same charge, for a given wormhole radius the BI source helps to reduce the amount of exotic matter (see Fig. 3(b)). The same result (less exotic matter in the BI case than in the Maxwell case) is obtained for values of the charge larger than unity (see Fig. 4). In other words, for a given charge, if it appears in a nontrivial manner, the exotic matter content of a wormhole configuration turns to be reduced when compared with the usual Maxwell case.

## V. STABILITY ANALYSIS

In this section we study the mechanical stability of the wormholes under small perturbations preserving the symmetry of the original configuration [12]. The dynamical evolution of the wormhole is determined by Eqs. (17) and (18), or by any of them and Eq. (19), and to complete the system we must add an equation of state that relates $p$ with $\sigma$, i.e., $p = p(\sigma)$. From Eq. (17) we have

$$\dot{a}^2 + g(a) = (2\pi a \sigma(a))^2. \tag{24}$$

We first note that the energy conservation equation can be written as

$$\dot{\sigma} = -2(\sigma + p)\frac{\dot{a}}{a}, \tag{25}$$

which can be integrated to give

$$\frac{a(\tau)}{a(\tau_0)} = \exp\left[-\frac{1}{2}\int_{\sigma_0}^{\sigma}\frac{d\sigma}{\sigma + p(\sigma)}\right]. \tag{26}$$

From Eq. (19), if the equation of state $p = p(\sigma)$ is given, then one can obtain $\sigma = \sigma(a)$.

Following the procedure introduced by Poisson and Visser, the analysis of the stability of the configuration can be reduced to the analogous problem of the stability of a particle in a one-dimensional potential $V(a)$ [12]. This is easy to see if we write Eq. (24) as

$$\dot{a}^2 = -V(a), \tag{27}$$

$$V(a) = g(a) - (2\pi a\sigma(a))^2. \tag{28}$$

Then, to study the stability we expand up to second order the potential $V(a)$ around the static solution $a_0$ (for which $\dot{a} = 0$, $\ddot{a} = 0$). For a stable configuration it is $V(a_0) = 0$ and $V'(a_0) = 0$ (where, as before, the prime means a derivative with respect to the radius). Then, Eq. (27) takes the following form:

$$\dot{a}^2 = -V''(a_0)(a - a_0)^2 + \mathcal{O}[(a - a_0)^3]. \tag{29}$$

To compute the derivatives it is convenient to define the parameter

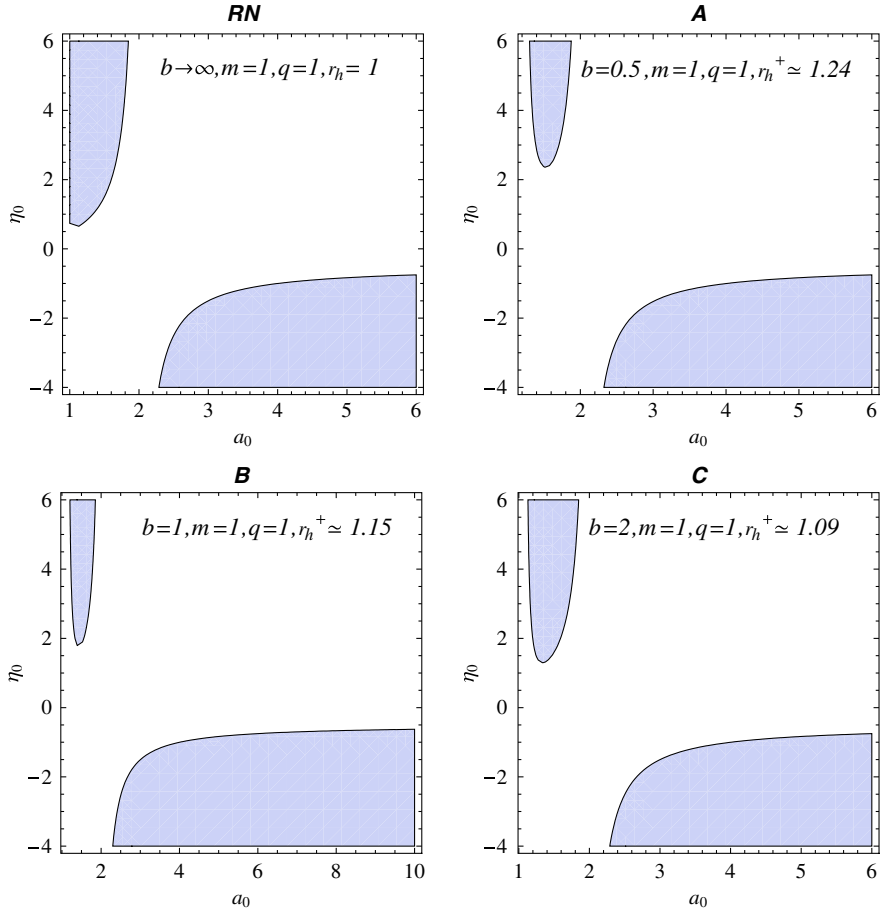

FIG. 5 (color online). We show the stability regions for $q = m = 1$ and four different values of the BI parameter.

$$\eta(\sigma) \equiv \frac{\partial p}{\partial \sigma}, \tag{30}$$

which for ordinary matter would represent the squared speed of sound: $v_s^2 = \eta$. Here, however, we simply consider $\eta$ as a parameter entering the equation of state (see below). Then, we obtain the second derivative of the potential for the metric (7)

$$V''(a_0) = g_0'' - \frac{g_0'^2}{2g_0} - \frac{1 + 2\eta_0}{a_0^2}[2g_0 - a_0 g_0'], \tag{31}$$

where we use the definitions $\eta_0 := \eta(\sigma_0)$, $g_0 := g(a_0)$, $g_0' := g'(a_0)$, and $g_0'' := g''(a_0)$ for short. The wormhole is stable if and only if $V''(a_0) > 0$, while for $V''(a_0) < 0$ a radial perturbation grows (at least until nonlinear regime is reached) and the wormhole is unstable. By using that the function $g(a_0)$ is always positive for $a_0 > r_{\text{hor}}$, we only have to analyze the sign of $V''(a_0)$ for determining which are the values of the parameters $(m, q, b, a_0)$ that make the wormhole stable. Then, after some simple manipulations, the stability conditions can be written as follows:

$$\text{if } 2g_0 > a_0 g_0', \qquad 1 + 2\eta_0 < \frac{a_0^2}{2g_0}\left(\frac{2g_0'' g_0 - g_0'^2}{2g_0 - a_0 g_0'}\right), \tag{32}$$

$$\text{if } 2g_0 < a_0 g_0', \; 1 + 2\eta_0 > \frac{a_0^2}{2g_0}\left(\frac{2g_0'' g_0 - g_0'^2}{2g_0 - a_0 g_0'}\right). \tag{33}$$

Note that the above expressions are in agreement with those found in [12,25]. Our starting point is to investigate how the regions of stability for the BI wormhole change with the parameter $b$. First, we check that when $b \to \infty$ the stability regions

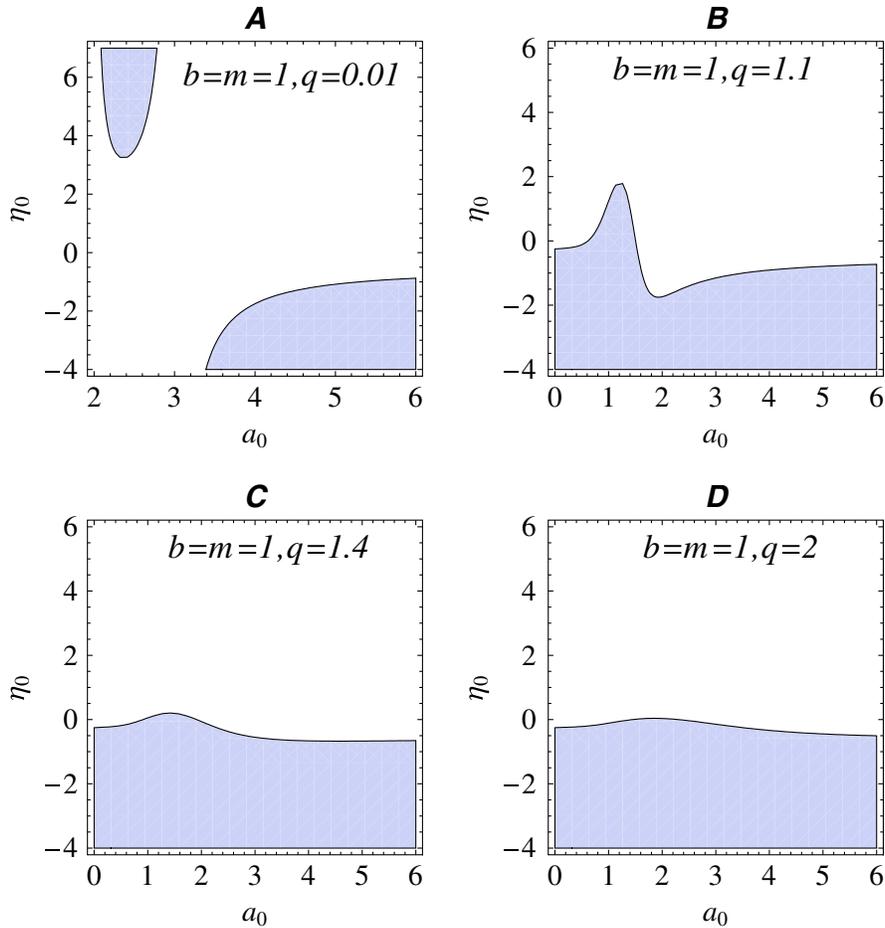

FIG. 6 (color online). We show the stability regions with $q_{\text{BI}} \in [1.1, 2]$ and $m = b = 1$. For comparison, we also exhibit the stability region for $q_{\text{BI}} = 0.01$.

correspond to the RN wormholes (see Fig. 5). By taking $b$ in the range [0.5, 2] we get that the stability regions remain almost without changes in relation to the extremal RN case. Then, we take the BI parameter as $b = 1$ (that is, considerably far away from the Maxwell limit) and consider the case $m = 1$. Then, we let the charge $q_{BI}$ vary in the interval [1.1, 2], so that the original manifold only presents a naked singularity at the origin (Fig. 6). In general, we obtain that there are different types of stability regions. Basically, we have a stable wormhole with $0 \leq \eta_0 \leq 1$ (indicating that $\eta_0$ could represent the speed of sound), $\eta_0 \leq 0$ (being this characteristic of wormhole matter), and a third case with $\eta_0 \geq 1$, which would correspond to a superluminal sound velocity on the shell placed at the wormhole throat. For $q_{BI} = 1.1$, the analysis shows that stable BI wormholes with $0 \leq \eta_0 \leq 1$ could exist. However, the same occurs within pure Maxwell theory. An interesting aspect is that for quite similar original manifolds one obtains very different stability domains when the BI parameter is $b = 1$. For example, in the case $q = 1.1$ (such that there is no horizon in the original manifold) stability can be achieved with $0 \leq \eta_0 \leq 1$; however, for $q = 1$ (when there is one horizon) stability requires $\eta_0$ larger than unity, or $\eta_0$ negative.

## VI. CONCLUSION

The generalization of Maxwell electromagnetism to a nonlinear theory in the way proposed by Born and Infeld introduces a new parameter, which allows for more freedom in the framework of determining the most viable charged wormhole configurations. If wormholes could actually exist, one would be interested in those that are stable—at least under the most simple kind of perturbations—and which, besides, require as little amount of exotic matter as possible. Of course, the case could be that a given change of the theory leads to a worse situation, i.e., that configurations turn out to be more unstable or require more matter violating the energy conditions as the departure from the standard theory becomes relevant. However, for large charges, this seems not to be the case with Born-Infeld electrodynamics coupled with Einstein's gravity: Here, we have examined the mechanical stability and exotic matter content of thin-shell wormholes within Einstein-Born-Infeld theory, and as long as large values of the charge are considered, we have found that for small values of the Born-Infeld parameter, corresponding to a situation far away from the Maxwell limit, the amount of exotic matter is reduced in relation to the standard case. Thus, if the requirement of exotic matter is considered as the strongest objection against wormholes, our results suggest that in a physical scenario different from that consistent with present day observation (as in the early Universe, when nonlinear effects could be more relevant) charged wormholes could have been more likely to exist.

### ACKNOWLEDGMENTS

We thank S. Habib Mazharimousavi and M. Halilsoy for their comment in which they pointed out the errors in the first version of the paper. M. G. R. and C. S. are supported by the Consejo Nacional de Investigaciones Científicas y Técnicas (CONICET).